\documentclass[aps,prd,twocolumn,nofootinbib,longbibliography,10pt]{revtex4-2}
\usepackage{etoolbox}
\usepackage{dcolumn}
\usepackage{amsmath,amssymb,amsfonts,mathtools}
\usepackage{mathrsfs}
\usepackage{graphicx}
\usepackage[colorlinks=true,urlcolor=blue,citecolor=red,linkcolor=blue]{hyperref}
\usepackage{natbib}
\usepackage{wrapfig}
\usepackage{flushend}

\makeatletter
\newsavebox{\@brx}
\newcommand{\llangle}[1][]{\savebox{\@brx}{\(\m@th{#1\langle}\)}%
  \mathopen{\copy\@brx\kern-0.5\wd\@brx\usebox{\@brx}}}
\newcommand{\rrangle}[1][]{\savebox{\@brx}{\(\m@th{#1\rangle}\)}%
  \mathclose{\copy\@brx\kern-0.5\wd\@brx\usebox{\@brx}}}
\makeatother
\begin{document}
\title{Uncertainty principle from the noise of gravitons}
\author{Soham Sen}
\email{sensohomhary@gmail.com}
\affiliation{Department of Astrophysics and High Energy Physics, S. N. Bose National Centre for Basic Sciences, JD Block, Sector-III, Salt Lake City, Kolkata-700 106, India}
\author{Sunandan Gangopadhyay}
\email{sunandan.gangopadhyay@gmail.com}
\affiliation{Department of Astrophysics and High Energy Physics, S. N. Bose National Centre for Basic Sciences, JD Block, Sector-III, Salt Lake City, Kolkata-700 106, India}
\begin{abstract}
\noindent The effect of the noise induced by gravitons in the case of a freely falling particle from the viewpoint of an external observer has been recently calculated in \href{https://link.aps.org/doi/10.1103/PhysRevD.107.066024}{Phys. Rev. D 107 (2023) 066024}. There the authors have calculated the quantum gravity modified Newton's law of free fall where the spacetime has been considered to be weakly curved. In our work, we extend this work by calculating the variance in the velocity and eventually the momentum of the freely falling massive particle. From this simple calculation, we observe that the product of the standard deviation in the position with that of the standard deviation in momentum picks up a higher order correction which is proportional to the square of the standard deviation in momentum. We also find out that in the Planck limit (both Planck length and Planck mass),  this uncertainty product gives the well-known form of the generalized uncertainty principle. We then calculate a similar uncertainty product when the graviton is in a squeezed state, and eventually, we get back the same uncertainty product. Finally, we  extend our analysis for the gravitons being in a thermal state and obtain a temperature-dependent uncertainty product. If one replaces this temperature with the Planck temperature and the mass of the particle by the Planck mass, the usual uncertainty product appears once again. We also obtain an upper bound of the uncertainty product thereby giving a range of the product of the variances in position and momentum. 
\end{abstract}
\maketitle
\section{Introduction}
\noindent The new age of theoretical physics involving graviton-particle interactions is directing towards a new area of high energy physics where deterministic classical equations are being dominated by stochastic quantum gravity fluctuations.  In a series of works \cite{QGravNoise,QGravLett,QGravD,Soda,OTMGraviton}, the interaction of a model interferometer detector and a gravitational wave has been explored. It has been observed that the geodesic deviation equation gets replaced by a Langevin-like equation which involves a stochastic noise term. Recently in \cite{AppleParikh}, the modification to Newton's law of free fall due to considering the effects of the quantum gravitational fluctuations on a freely falling massive particle, has been discussed. Instead of considering a flat spacetime geometry, the authors in \cite{AppleParikh} have considered a slightly curved background geometry where the observer is fixed with respect to the background. The idea used in \cite{AppleParikh} is to use perturbation on the slightly curved geometry for which the post-Newtonian modified spacetime metric is used. Finally, the fluctuation over the background metric has been quantized to directly involve quantum gravitational calculation in the entire analysis. The calculation shows that apart from the special and general relativistic corrections, Newton's universal law of gravitation now contains a stochastic noise term due to the quantum gravity correction introduced during the calculation. In a recent analysis \cite{Giulia}, formulation of the quantum gravity-matter system in a WKB approach has been proposed. 

\noindent In our work, we have extended the calculation in \cite{AppleParikh} and obtained an uncertainty relation induced from the noise of gravitons. Contrary to the calculation done in \cite{AppleParikh}, we have restricted our calculation to Newtonian approximation only. It is important to note that the leading order term for the variance in the position value has already been calculated in \cite{AppleParikh} for a coherent state and later argued for a squeezed state. We have calculated the corresponding variance in the momentum for the freely falling particle. We find out that the product of the variances in the position and momentum variables for the graviton being in a coherent state is proportional to the square of the variance in the momentum. We then calculate the uncertainty product for the graviton being initially in a squeezed state. Instead of the individual values of the standard deviation of the position and momentum variables being amplified by the squeezing parameter, we find out that the product remains the same as in the previous case. We find out that the minimum value of the uncertainty product now depends on the mass of the particle and it depends on both Planck's constant and Newton's gravitational constant. To our surprise, we find out that in the limit of the particle mass being replaced by the Planck mass, the uncertainty principle induced by the noise of gravitons now exactly resembles the usual generalized uncertainty principle with quadric correction in the momentum uncertainty as introduced by Kempf \textit{et. al.} in \cite{Kempf} and later used in various literatures \cite{MAGGIORE,SCARDIGLI,ADLERSAN,
ADLERCHENSAN,RABIN,SG1,SCARDIGLI2,SG2,Ong,EPJC,BMajumder,DAS1,DAS2,
IVASP,IVASP2,KSP,SG3,SG4,OTM,Petruzzeillo,
DasModak,Farag,Farag2,SGSB,OTM0,Vagnozzi,Feng,HCuletu} (with linear order corrections in the standard deviation in the momentum as well). It is important to note that if a particle is more massive than Planck's mass then its Compton wavelength gets smaller than the Planck length hence writing down this relation for a particle that is more massive than Planck's mass is not very useful. If one carefully looks at this uncertainty relation along with all its limiting values, it is straightforward to observe that the form of the generalized uncertainty relation used usually in the literature serves as the smallest possible value of the uncertainty product for such a quantum gravity calculation. We have finally considered the graviton to be initially in a thermal state. The thermal state calculation gives an uncertainty product which now depends on Planck's constant, Boltzmann constant, temperature of the thermal gravitons, and mass of the particle. It is very interesting to see that the coefficient of the square of the variance of the momentum reduces to the earlier case if one replaces the temperature with the Planck temperature. If one again resubstitutes the mass of the particle with the Planck mass, we get back the uncertainty product corresponding to the well-known generalized uncertainty principle. Another important aspect that has been revealed in this calculation is that the uncertainty product always has a upper bound.

\noindent Our primary motivation behind this work is to investigate the form of the uncertainty principle when true quantum gravity effects are being considered. It is easy to see that this is one of the most fundamental setups (Newtonian free fall) one can consider while doing a quantum gravity calculation. As the entire calculation is done using Newtonian approximation, it is straightforward to state that the uncertainty relation obtained in this case will break Lorentz invariance. We have got rid of the post-Newtonian correction as they play no important role in the stochastic part of the equation of motion. The most important aspect of this new uncertainty relation is that the minimum value of the uncertainty product now depends on (apart from the square of the variance of the momentum) Planck's constant and Newton's gravitational constant which is not present in the case of the well known generalized uncertainty principle with quadratic momentum term. This signifies that this uncertainty relation indicates a true quantum gravitational effect rather than a pseudo one. 

\noindent Our paper is organized as follows. In section \ref{S2}, we briefly review the model and the results given in \cite{AppleParikh}. In section \ref{S3}, we obtain the product of the variance for the position and momentum variables, and summing with the general uncertainty product value ($\frac{\hbar}{2}$) we obtain the modified uncertainty relation. We also repeat the same calculation for the gravitational fluctuation being in a squeezed vacuum state and eventually a thermal state. Finally, we summarize our results in section \ref{S4} and conclude.

\section{Brief review of the model} \label{S2}
\noindent In this section, we briefly review the underlying model in \cite{AppleParikh} and discuss some of the important results obtained in this analysis.  The line element for the background metric is given as follows
\begin{equation}\label{1.1}
\begin{split}
ds^2=&-(1+2\phi+2\phi^2+2\psi+\cdots)dt^2+(\delta_{ij}-2\phi\delta_{ij}\\&+g_{ij}^{(4)}+\cdots)dx^idx^j
\end{split}
\end{equation}
where $\phi,\psi$ and $g_{ij}^{(4)}$ are leading order post Newtonian corrections \cite{Weinberg}. The idea is to consider small fluctuations ($h_{\mu\nu}$) over the background metric and following the arguments presented in \cite{AppleParikh}, one can write down the fluctuation in the transverse traceless gauge. The action describing the motion of the freely falling point particle of mass $m_0$ reads \cite{AppleParikh}
\begin{equation}\label{1.2}
\begin{split}
S_p=&m_0\int dt\Bigr(\frac{1}{2}\dot{\xi}^2+\frac{1}{8}\dot{\xi}^4-\frac{3}{2}\phi\dot{\xi}^2-\phi-\frac{1}{2}\phi^2-\psi\\&+\frac{1}{4}\ddot{h}_{jk}\xi^j\xi^k\Bigr)
\end{split}
\end{equation}
where $\xi$ denotes the coordinate of the freely falling particle. 

\noindent The Einstein-Hilbert action on the other hand can be written as
\begin{equation}\label{1.3}
S_{EH}=-\frac{1}{64\pi G}\int d^4x\partial_\kappa h_{ij}\partial^\kappa h^{ij}~.
\end{equation}
One can now make use of a discrete mode decomposition of the fluctuation term $h_{ij}$ given as \cite{QGravD} (in a finite volume $V$)
\begin{equation}\label{1.4}
h_{ij}(t,\vec{x})=\frac{1}{\sqrt{\hbar G}}\sum_{\vec{k},s}q_{\vec{k},s}e^{i\vec{k}\cdot\vec{x}}\varepsilon^s_{ij}(\vec{k})
\end{equation}
with $\varepsilon^s_{ij}(\vec{k})$ denoting the polarization tensor of the gravitational fluctuation. Restricting the system to the $`+'$ polarization of the gravitational fluctuation, a single mode of frequency $\omega$ and one spatial dimension, one can write down the Lagrangian for the system in consideration to be
\begin{equation}\label{1.5}
\begin{split}
L=&m_0\left(\frac{1}{2}\dot{z}^2+\frac{1}{8}\dot{z}^4-\frac{3}{2}\phi\dot{z}^2-\phi-\frac{1}{2}\phi^2-\psi\right)+\\
&+\frac{1}{2}m\left(\dot{q}^2-\omega^2q^2\right)-\mathcal{G}\dot{q}\dot{z}z
\end{split}
\end{equation}
where $m=\frac{V}{16\pi\hbar G^2}$,  $\mathcal{G}=\frac{m_0}{2\sqrt{\hbar G}}$, and $\xi_1=z$.  From the above Lagrangian, it is straightforward to write down the Hamiltonian of the system given by \cite{AppleParikh}
\begin{equation}\label{1.6}
H\simeq\frac{\frac{p^2}{2m}+\frac{\frac{\pi^2}{2}+\frac{\mathcal{G}p\pi z}{m}}{m_0(1-3\phi)}}{1-\frac{\mathcal{G}^2z^2}{mm_0(1-3\phi)}}+\frac{1}{2}m\omega^2q^2+m_0\left(\phi+\frac{\phi^2}{2}+\psi\right)
\end{equation}
where $\phi=\phi(z)$ and $\psi=\psi(z)$ with $\pi$ denoting the conjugate momentum of $z$, and $p$ denoting the conjugate momentum of $q$. The next step is to quantize the system by raising the position and momentum variables to operator status. If the graviton is initially in a state $|\Psi\rangle$ with the freely falling particle being in a state $|A\rangle$ and finally the state of the graviton is $|f\rangle$ with the particle state being denoted by $|B\rangle$ then the probability of transition reads
\begin{equation}\label{1.7}
P^{\Psi}_{A\rightarrow B}=\sum_{|f\rangle}\left|\langle f,B|\hat{U}(T,0)|\Psi,A\rangle\right|^2
\end{equation}
with $T$ denoting the particle-graviton interaction time. Using a path integral approach along with following the method in \cite{QGravNoise,QGravLett,QGravD,AppleParikh}, and making use of the Feynman-Vernon trick \cite{FeynmanVernon}, one can obtain the final form of the transition probability to be \cite{AppleParikh}
\begin{equation}\label{1.8}
\begin{split}
&P^{\Psi}_{A\rightarrow B}=\int dz_i dz_i' dz_f dz_f' \phi_A^{*}(z_i')\phi_A(z_i)\phi_B^{*}(z_f)\phi_B(z_f')\\
&\times\int [\tilde{\mathcal{D}}z]_{z_i,0}^{z_f,T} [\tilde{\mathcal{D}}z']_{z_i',0}^{z_f',T}\int \mathcal{D}\mathcal{N} \exp\Bigr[-\frac{1}{2}\int_0^Tdt\int_0^Tdt'\\&\times \mathcal{A}_{\Psi}^{-1}(t,t')\mathcal{N}(t)\mathcal{N}(t')\Bigr]\exp\Bigr[\frac{im_0}{\hbar}\int_0^Tdt\Bigr(\left(\mathcal{I}(z)-\mathcal{I}(z')\right)\\&+\frac{1}{4}\mathcal{N}(t)(X(t)-X'(t))-\frac{m_0G}{8}(X(t)-X'(t))(\dot{X}(t)\\&+\dot{X}'(t))\Bigr)\Bigr]
\end{split}
\end{equation}
where $X(t)=\frac{d^2}{dt^2}(z^2(t))$, $\mathcal{N}(t)$ is a random fluctuation term with $\llangle\mathcal{N}(t)\rrangle=0$, $\mathcal{A}_\Psi(t,t')=\llangle \mathcal{N}(t)\mathcal{N}(t')\rrangle$, and 
\begin{equation}\label{1.9}
\mathcal{I}(z)=\frac{1}{2}\dot{z}^2-\frac{3}{2}\phi(z)\dot{z}^2-\phi(z)-\frac{1}{2}\phi(z)^2-\psi(z)~.
\end{equation}
Using the saddle point approximation, one can find out the corresponding Langevin-like equation for $z$ and neglecting all other corrections except the stochastic term, one can write down Newton's law as 
\begin{equation}\label{1.10}
F=m_0\ddot{z}(t)\simeq m_0\left(-g+\frac{1}{2}\ddot{\mathcal{N}}(t)z(t)\right)
\end{equation}
which is the same as the result obtained in \cite{AppleParikh} up to a half factor in front of the noise term.
Here, $g$ is the acceleration due to gravitation on Earth. 
\section{New uncertainty relation from graviton noise}
\label{S3}
\noindent We shall again use the result of the square of the variance of the position from \cite{AppleParikh}.  The trajectory of the particle which is freely falling under the acceleration of Earth is $z(t)=z_0-\frac{1}{2}gt^2$. If at time $t=\tau$ the particle hits the ground then $z(\tau)=0$ when we are only considering the classical part of $z(t)$. As the Euler-Lagrange equation now depends on a noise part due to the interaction of the gravitons with the particles, it is imperative to break $z(t)$ into a classical part and a quantum part as 
\begin{equation}\label{1.11}
z(t)\simeq z_{cl}(t)+z_{q}(t)~.
\end{equation}
Substituting the above result back in eq.(\ref{1.10}), it is straight forward to see that
\begin{equation}\label{1.12}
\ddot{z}_q(t)\simeq \frac{1}{2}\ddot{\mathcal{N}}(t)z_{cl}(t).
\end{equation}
Integrating only once, we get back the form of the first time derivative of $z_q(t)$ as follows
\begin{equation}\label{1.13}
\begin{split}
\dot{z}_q(t)&\simeq\frac{1}{2}\dot{\mathcal{N}}(t)z_{cl}(t)-\frac{1}{2}\mathcal{N}(t)\dot{z}_{cl}(t)+\frac{1}{2}\int_0^tdt'\mathcal{N}(t')\ddot{z}_{cl}(t')\\
&=\frac{1}{2}\dot{\mathcal{N}}(t)z_{cl}(t)+\frac{1}{2}gt\mathcal{N}(t)-\frac{g}{2}\int_0^tdt'\mathcal{N}(t')~.
\end{split}
\end{equation}
Integrating eq.(\ref{1.13}) once again, one can obtain the form of $z_q(t)$ as follows
\begin{equation}\label{1.14}
\begin{split}
z_q(t)&\simeq\frac{1}{2}\mathcal{N}(t)z_{cl}(t)+\int_0^tdt' gt'\mathcal{N}(t') \\&-\frac{1}{2}\int_0^tdt'\int_0^{t'} dt''g\mathcal{N}(t'')
\end{split}
\end{equation}
which is same as the result in \cite{AppleParikh} upto an overall $\frac{1}{2}$ factor. As $\mathcal{N}(t)$ is a stochastic term, therefore it is straightforward to conclude that 
\begin{equation}\label{1.15}
\llangle z_q(t)\rrangle=0~.
\end{equation}
Using eq.(\ref{1.15}), it is possible to write down the following relation
\begin{equation}\label{1.16}
\llangle z(t)\rrangle=\llangle (z_{cl}(t)+z_q(t))\rrangle=z_{cl}(t).
\end{equation}
The goal is to measure variance in $z(t)$ at the time of the particle touching the ground. Hence, at $t=\tau$ the classical part of $z(t=\tau)$ must vanish which indicates $z_{cl}(\tau)=0$. The variance in position (at time $t=\tau$) is given as \cite{AppleParikh}
\begin{equation}\label{1.17}
\begin{split}
(\Delta z)^2&=\llangle z(\tau) z(\tau)\rrangle-\llangle z(\tau)\rrangle^2\\
&=\llangle z(\tau) z(\tau)\rrangle-z_{cl}(\tau)^2\\
&=\llangle z_q(\tau) z_q(\tau)\rrangle\\
&=g^2\int_0^\tau t' dt'\int_0^\tau \bar{t} d\bar{t} \llangle \mathcal{N}(t')\mathcal{N}(\bar{t})\rrangle\\
&-g^2\int_{0}^\tau dt'\int_0^{t'} dt''\int_0^{\tau}\bar{t}d\bar{t}\llangle \mathcal{N}(t'')\mathcal{N}(\bar{t})\rrangle\\&
+\frac{g^2}{4}\int_0^{\tau}dt'\int_0^{t'}dt''\int_{0}^\tau d\bar{t}\int_0^{\bar{t}}d\bar{\bar{t}}\llangle\mathcal{N}(t'')\mathcal{N}(\bar{\bar{t}})\rrangle~.
\end{split}
\end{equation}
As we are mainly focused on the noise part of the solution, we can get rid of all of the post-Newtonian correction terms in the Lagrangian in eq.(\ref{1.5}). Hence, in the Newtonian approximation only, we can recast eq.(\ref{1.5}) in the following form
\begin{equation}\label{1.18}
\begin{split}
L=&\frac{1}{2}m_0\dot{z}^2-m_0\phi(z)+\frac{1}{2}m\left(\dot{q}^2-\omega^2q^2\right)-\mathcal{G}\dot{q}\dot{z}z
\end{split}
\end{equation}
where 
\begin{equation}\label{1.19}
\pi_z=\frac{\partial L}{\partial \dot{z}}=m_0\dot{z}-\mathcal{G}\dot{q}z~.
\end{equation}
We are interested in calculating $(\Delta \pi_z)^2$ when $t=\tau$. Using eq.(\ref{1.11}), we can recast $p_z(\tau)$ in the following form
\begin{equation}\label{1.20}
\begin{split}
\pi_z(\tau)&=m_0(\dot{z}_{cl}(\tau)+\dot{z}_q(\tau))-\mathcal{G}\dot{q}(\tau)\left(z_{cl}(\tau)+z_q(\tau)\right)\\
&=m_0(\dot{z}_{cl}(\tau)+\dot{z}_q(\tau))-\mathcal{G}\dot{q}(\tau)z_q(\tau)~.
\end{split}
\end{equation}
It is important to note that $\mathcal{G}$ is a coupling constant multiplied to $z_q(\tau)$ which is purely a quantum mechanical term and as a result the total term becomes very small compared to the two terms preceding it. We can therefore,  upto a good approximation, write down
\begin{equation}\label{1.21}
\pi_z(\tau)\simeq m_0(\dot{z}_{cl}(\tau)+\dot{z}_q(\tau))~.
\end{equation}
Hence, one can compute the square of the variance in the momentum variable to be
\begin{equation}\label{1.22}
\begin{split}
(\Delta\pi_z(\tau))^2=&m_0^2\left(\llangle\dot{z}(\tau)\dot{z}(\tau)\rrangle-\llangle\dot{z}(\tau)\rrangle^2\right)\\
=&m_0^2(\dot{z}_{cl}^2(\tau)+2\dot{z}_{cl}(\tau)\llangle \dot{z}_q(\tau)\rrangle+\llangle \dot{z}_q(\tau)\dot{z}_q(\tau)\rrangle\\&-\dot{z}_{cl}^2(\tau))\\
=&m_0^2(2\dot{z}_{cl}(\tau)\llangle \dot{z}_q(\tau)\rrangle+\llangle \dot{z}_q(\tau)\dot{z}_q(\tau)\rrangle)~.
\end{split}
\end{equation}
From eq.(\ref{1.13}) it is straighforward to infer that $\llangle \dot{z}_q(\tau)\rrangle=0$. Using this relation, we can recast eq.(\ref{1.22}) in the following form
\begin{equation}\label{1.23}
(\Delta\pi_z(\tau))^2=m_0^2\llangle\dot{z}_q(\tau)\dot{z}_q(\tau)\rrangle~.
\end{equation}
We shall now calculate the uncertainty product in case of the graviton being in a vacuum, squeezed vacuum state, and a thermal state. 
\subsection{Vacuum state}
\noindent We start with the simple consideration of the graviton being in a vacuum state. In that case, the two-point correlator corresponding to the stochastic term ($\mathcal{N}(t)+\mathcal{N}_0(t)$) takes the form
\begin{equation}\label{1.24}
\begin{split}
\llangle \mathcal{N}_0(t')\mathcal{N}_0(t'')\rrangle&=\mathcal{A}_{0}(t',t'')=\int_0^\infty d\omega \omega \cos(\omega(t'-t''))~.
\end{split}
\end{equation}
From eq.(\ref{1.24}), it is evident that the integral over $\omega$ is divergent. The way to deal with this problem is to regularize the upper bound of the integral by some finite value of the frequency $\omega=\omega_{\text{max}}=\frac{2\pi c}{z_0}$, where $z_0$ is the initial height of the particle from the surface of the Earth. There is also a minimum value of the frequency term as the measurement is done over a finite amount of time $t=\tau$. Hence, the lower limit of integration is $\omega_{\text{min}}=\frac{2\pi}{\tau}$. Using the limits of integration, we can integrate eq.(\ref{1.24}) to get
\begin{equation}\label{1.25}
\begin{split}
&\mathcal{A}_0(t',t'')\\&=\frac{4\hbar G}{\pi c^5}\int_{\omega_{\text{min}}}^{\omega_{\text{max}}} d\omega \omega \cos\left(\omega (t'-t'')\right)\\
&=\frac{4\hbar G}{\pi c^5}\left(\frac{\cos(\omega(t'-t''))}{(t-t')^2}+\frac{\omega\sin(\omega(t'-t''))}{t'-t''}\right)\biggr\rvert_{\omega_{\text{min}}}^{\omega_{\text{max}}}~.
\end{split}
\end{equation}
In case of the graviton initially being in a vacuum state, we can recast eq.(\ref{1.17}) in the following form
\begin{equation}\label{1.26}
\begin{split}
(\Delta z)^2&=\llangle z_q(\tau) z_q(\tau)\rrangle\\
&=g^2\int_0^\tau t' dt'\int_0^\tau \bar{t} d\bar{t} \mathcal{A}_0(t',\bar{t})\\
&-g^2\int_{0}^\tau dt'\int_0^{t'} dt''\int_0^{\tau}\bar{t}d\bar{t}\mathcal{A}_0(t'',\bar{t})\\&
+\frac{g^2}{4}\int_0^{\tau}dt'\int_0^{t'}dt''\int_{0}^\tau d\bar{t}\int_0^{\bar{t}}d\bar{\bar{t}}\mathcal{A}_0(t'',\bar{\bar{t}})~.
\end{split}
\end{equation} 
Using eq.(\ref{1.25}) back in eq.(\ref{1.26}), we can obtain the following relation
\begin{equation}\label{1.27}
\begin{split}
(\Delta z)^2&\simeq \frac{4\hbar g^2 G}{\pi c^5}\biggr[\frac{9z_0^2}{16\pi^2c^2}\left(\cos\left(\frac{2\pi c\tau}{z_0}\right)-1\right)\\&+\frac{1}{2}\tau^2\left(\text{Ci}(2\pi)-\text{Ci}\left(\frac{2\pi c\tau}{z_0}\right)\right)\\&+\frac{1}{8}\tau^2\ln\left(\frac{4\pi^2}{\tau^2}\right)+\tau^2\ln\left(\frac{c\tau}{z_0}\right)\biggr]\\
\implies (\Delta z)^2&\simeq \frac{4\hbar g^2 G\tau^2}{\pi c^5}\ln\left(\frac{c\tau}{z_0}\right)=\frac{4g^2\tau^2l_p^2}{\pi c^2}\ln\left(\frac{c\tau}{z_0}\right)
\end{split}
\end{equation}
where in the last line of the above equation we have kept only the logarithmic term as it dominates with increasing values of the parameter $\tau$. In the above equation, `$\text{Ci}$' denotes the cosine integral function \cite{Gradshteyn}.The result is similar to the result that has been obtained in \cite{AppleParikh} up to a constant factor. For the next part of our calculation, we shall calculate the variance in the momentum $(\Delta \pi_z)^2$
\begin{equation}\label{1.28}
\begin{split}
(\Delta \pi_z)^2=&\frac{2m_0^2g^2l_p^2}{\pi c^2}\Biggr[\frac{\pi^2c^2\tau^2}{z_0^2}+\left(\cos\left(\frac{2\pi c\tau}{z_0}\right)-1\right)\\-&\pi^2+\text{Ci}(2\pi)-\text{Ci}\left(\frac{2\pi c\tau}{z_0}\right)+\ln\left(\frac{c\tau}{z_0}\right)\Biggr]\\
\implies (\Delta \pi_z)^2\simeq&\frac{2\pi m_0^2g^2l_p^2\tau^2}{z_0^2}
\end{split}~.
\end{equation}
In the last line of the above equation, we have again kept the dominant contribution to the square of the variance only. Multiplying eq.(\ref{1.27}) with eq.(\ref{1.28}), we obtain 
\begin{equation}\label{1.29}
\begin{split}
(\Delta z)^2(\Delta \pi_z)^2=\frac{8m_0^2g^4\tau^4l_p^4}{z_0^2c^2}\ln\left(\frac{c\tau}{z_0}\right)\\
\implies \Delta z\Delta\pi_z=\frac{2\sqrt{2}m_0g^2\tau^2l_p^2}{z_0 c}\sqrt{\ln \left(\frac{c\tau}{z_0}\right)}~.
\end{split}
\end{equation}
It is important to note that the right-hand side can be represented in terms of the square of the variance in the momentum parameter as follows
\begin{equation}\label{1.30}
\Delta z\Delta\pi_z=\frac{2\pi m_0^2g^2\tau^2l_p^2}{z_0 c}\frac{z_0}{m_0 c}\sqrt{\frac{2}{\pi^2}\ln\left(\frac{c\tau}{z_0}\right)}=\frac{\beta z_0}{m_0 c}(\Delta \pi_z)^2
\end{equation}
where in the above equation $\beta\equiv \sqrt{\frac{2}{\pi^2}\ln\left(\frac{c\tau}{z_0}\right)}=\sqrt{\frac{1}{\pi^2}\ln\left(\frac{2c^2}{gz_0}\right)}$ where $\tau=\sqrt{\frac{2z_0}{g}}$.
From eq.(\ref{1.30}), we can obtain a lower bound on the uncertainty product for a particle with mass $m_0$ as
\begin{equation}\label{1.31}
\Delta z\Delta \pi_z=\frac{\beta z_0}{m_0 c}(\Delta \pi_z)^2\geq\frac{\beta l_p}{m_0 c}(\Delta \pi_z)^2~.
\end{equation}
To write down the above inequality, we have considered the lowest value possible that $z_0$ can pick up for a quantum gravity calculation which is the Planck length $l_p=\sqrt{\frac{\hbar G}{c^3}}$. 
Now $\Delta z\Delta\pi_z$ must obey the usual Heisenberg uncertainty product while considering a quantum detector model as
\begin{equation}\label{1.32}
\Delta z\Delta \pi_z\geq \frac{\hbar}{2}~.
\end{equation}
Using the same argument as used in \cite{Scardigli}, we combine eq.(\ref{1.31}) and eq.(\ref{1.32}) and write down the total uncertainty product as 
\begin{align}
\Delta z\Delta \pi_z&\geq \frac{\hbar}{2}+\frac{\beta l_p}{m_0c}(\Delta \pi_z)^2\nonumber\\
&=\frac{\hbar}{2}\left(1+\frac{2\beta l_p}{\hbar m_0 c}(\Delta \pi_z)^2\right)\label{1.33}\\
\implies\Delta z\Delta \pi_z&\geq \frac{\hbar}{2}+\frac{\beta l_p}{m_0c}(\Delta \pi_z)^2\nonumber\\&\geq\frac{\hbar}{2}+\frac{\beta l_p}{m_pc}(\Delta \pi_z)^2\nonumber\\
&=\frac{\hbar}{2}+\frac{\beta G}{c^3}(\Delta\pi_z)^2\nonumber\\
\implies \Delta z\Delta\pi_z &\geq \frac{\hbar}{2}\left(1+\frac{\beta'}{m_p^2c^2}(\Delta\pi_z)^2\right)\label{1.34}
\end{align}
where $\beta'=2\beta$. It is important to note from eq.(\ref{1.33}) that the correction to Heisenberg's uncertainty principle now depends on three fundamental constants, $\hbar$, $G$, and the speed of light $c$. For the well-known generalized uncertainty relation the coefficient of the square of the variance of the momentum depends solely on $G$ and $c$ and there is no involvement of the Planck's constant. We can recover the generalized uncertainty principle with quadratic order momentum correction from eq.(\ref{1.33}) by substituting $m_0$ by $m_p=\sqrt{\frac{\hbar c}{G}}$. This limit is also important in a quantum gravitational scenario as it describes the maximum mass one can confine inside the Planck volume. If the mass exceeds the Planck mass, it would turn into a quantum black hole. It is important to note that, we are indeed considering the freely falling particle to be of mass much smaller than Planck's mass. Hence, throughout our analysis $m_0<m_p$. We obtain the final form of the uncertainty relation in eq.(\ref{1.34}) by substituting $m_0$ with the Planck mass $m_p$.
Eq.(\ref{1.34}) resembles the uncertainty product as given in \cite{Kempf} for one dimension. This can also be interpreted as a derivation of the well-known generalized uncertainty principle form. As argued earlier, the coefficient of the $(\Delta\pi_z)^2$ term in eq.(\ref{1.34}) has no dependence on Planck's constant. It is unusual in the sense that any quantum gravity correction must involve both Planck's constant and Newton's gravitational constant which is present in the new uncertainty product obtained in eq.(\ref{1.33}). From the equality condition in eq.(\ref{1.33}), we find out that 
\begin{equation}\label{1.00a}
\Delta\pi_z=\frac{m_0 c}{\beta'l_p}\Delta z\pm \frac{m_0 c}{\beta'l_p}\sqrt{(\Delta z)^2-\frac{\beta'\hbar l_p}{m_0c}}~.
\end{equation}
From the above relation it is straightforward to interpret that
\begin{equation}\label{1.01b}
\begin{split}
\Delta z&\geq \sqrt{\frac{\beta'\hbar l_p}{m_0c}}\implies \Delta z_{\text{min}}= \sqrt{\frac{\beta'\hbar l_p}{m_0c}}~.
\end{split}
\end{equation}
In the $m_0\rightarrow m_p$ limit, the value of the minimum value for the position uncertainty reduces to the value $\Delta z_{\text{min}}=\sqrt{\frac{\hbar\beta'l_p}{m_p c}}=l_p\sqrt{\beta'}$ which is identical to the minimum value of the position uncertainty obtained in \cite{Kempf}. The above analysis suggests that $\Delta z_{min}\rvert_{m_0\rightarrow m_p}\leq \Delta z_{\text{min}}$ for all $m_0\leq m_p$. This is fine considering the fact that the minimum observable length under no circumstaces should have a value smaller than the Planck length in a quantum gravitational scenario.

\noindent A very non-intuitive bound can be obtained from eq.(\ref{1.30}) on the uncertainty product if the uncertainty product obtained in eq.(\ref{1.29}) is expressed in terms of the variance in the position part. We can recast eq.(\ref{1.29}) in the following form
\begin{equation}\label{1.35}
\begin{split}
\Delta z\Delta\pi_z&=\frac{2\sqrt{2}m_0g^2\tau^2l_p^2}{z_0 c}\sqrt{\ln \left(\frac{c\tau}{z_0}\right)}\\&=\sqrt{\frac{\pi^2}{2\ln\left(\frac{c\tau}{z_0}\right)}}\frac{m_0c}{z_0}\frac{4g^2\tau^2l_p^2}{\pi c^2}\ln\left(\frac{c\tau}{z_0}\right)\\
&=\frac{m_0 c}{\beta z_0}(\Delta z)^2~.
\end{split}
\end{equation}
If we now, replace $z_0$ by $l_p$, we get an upper bound to the uncertainty product given as
\begin{equation}\label{1.36}
\Delta z\Delta\pi_z=\frac{m_0 c}{\beta z_0}(\Delta z)^2\leq \frac{m_0 c}{\beta l_p}(\Delta z)^2~.
\end{equation}
As we are working with particles that have masses smaller than the Planck mass, we can get an even higher value of the upper bound simply by replacing $m_0$ with $m_p$ in the above inequality.
Hence, combining eq.(\ref{1.36}) with eq.(\ref{1.33}), we can define a range for the uncertainty product given as
\begin{equation}\label{1.37}
\frac{m_p c}{\beta l_p}(\Delta z)^2>\Delta z\Delta \pi_z\geq \frac{\hbar}{2}\left(1+\frac{\beta' l_p}{\hbar m_0 c}(\Delta \pi_z)^2\right)~.
\end{equation}
The above result has a very serious significance. It states that due to the inclusion of a stochastic noise term, the true quantum gravitational effects restrict one from precisely measuring the position of the particle. The value of $\Delta z$ will always have a non-zero value. $\Delta z$ can atmost be of the order of the Planck length $l_p$. Although the same cannot be claimed for the measurement of the momentum parameter. As the upper bound depends on $(\Delta z)^2$, one can always have a precise measurement of $\pi_z$ which will indicate an infinite value of $\Delta z$ and as a result, the upper bound criteria will still be satisfied.
We shall investigate a bit more into this upperbound criteria. From eq.(\ref{1.01b}), we have obtained
\begin{equation}\label{1.02c}
\Delta \pi_z=\frac{m_0 c}{\beta' l_p}\Delta z\pm\frac{m_0 c}{\beta' l_p}\sqrt{(\Delta z)^2-(\Delta z_{\text{min}})^2}
\end{equation}
 Multiplying the above equation with $\Delta z$, we get back from eq.(\ref{1.02c})
\begin{equation}\label{1.03d}
\Delta z\Delta\pi_z=\frac{m_0 c}{\beta'l_p}(\Delta z)^2\left(1\pm\sqrt{1-\frac{(\Delta z_{\text{min}})^2}{(\Delta z)^2}}\right)~.
\end{equation}
From eq.(\ref{1.37}), we have the value of the upperbound to be $\frac{2m_p c}{\beta'l_p}(\Delta z)^2$. As, $m_p \geq m_0$ and $2>\left(1\pm\sqrt{1-\frac{(\Delta z_{\text{min}})^2}{(\Delta z)^2}}\right)$, we obtain
\begin{equation}\label{1.04e}
\frac{2m_p c}{\beta'l_p}(\Delta z)^2>\frac{m_0 c}{\beta'l_p}(\Delta z)^2\left(1\pm\sqrt{1-\frac{(\Delta z_{\text{min}})^2}{(\Delta z)^2}}\right)~.
\end{equation}
The above inequality satisfies the upperbound criteria obtained in eq.(\ref{1.37}). For $\Delta z=\Delta z_{\text{min}}$, we obtain from eq.(\ref{1.02c}), $\Delta \pi_z \Delta z=\sqrt{\frac{\hbar m_0 c}{\beta' l_p}}\Delta z_{\text{min}}=\hbar$. This is again found out to be smaller than the upper bound value $\frac{2\hbar m_p}{m_0}$. This indicates that for the most precise measurement of $z$ parameter possible ($\Delta z=\Delta z_{\text{min}}$), the maximum uncertainty in the measurement of $\pi_z$ will be $\Delta{\pi_z}_{\text{max}}=2m_p\sqrt{\frac{\hbar c}{\beta'm_0 l_p}}$. In the next two subsections, we shall be mainly focussed about the universality of the lower bound of the uncertainty relation induced by the noise of the gravitons in a quantum gravitational set up.
 We shall now proceed to calculate the uncertainty product when the graviton is initially in a squeezed state.
 
\subsection{Squeezed vacua}
\noindent For the gravitational fluctuation to be in a squeezed state, one needs to write the initial graviton state to be in the form $\hat{S}({\zeta_\omega})|0_\omega\rangle$ with the form of the squeezing operator given by $\hat{S}({\zeta_\omega})=e^{\frac{1}{2}\zeta_\omega^*\hat{a}^2-\zeta_\omega\hat{a}^{\dagger2}}$ with the complex squeezing parameter defined as $\zeta_\omega=r_\omega e^{i\phi_\omega}$. Following the analysis presented in \cite{QGravNoise,QGravLett,QGravD,OTMGraviton}, we can write down the transition probability in case of the graviton being initially in a squeezed state as
\begin{equation}\label{1.38}
\begin{split}
&P^\Psi_{A\rightarrow B}=\int dz_i dz_i' dz_fdz_f'\phi^*_A(z_i')\phi_A(z_i)\phi_B^*(z_f)\phi_B(z_f')\\\times&
\int \tilde{\mathcal{D}}z\tilde{\mathcal{D}}z' e^{\frac{i m_0}{\hbar}\int_0^T dt\left(\frac{1}{2}(\dot{z}^2-\dot{z}^{'2})-(\phi(z)-\phi(z'))\right) }F_{\zeta_\omega}[z,z']
\end{split}
\end{equation}
where $F_{\zeta_\omega}[z,z']=F_{0_\omega}[z,z']e^{i\Phi_{\zeta_\omega}[z,z']}$ is the total influence functional. In eq.(\ref{1.38}), $z(0)=z_i$, $z'(0)=z_i'$, $z(T)=z_f$, and $z'(T)=z_f'$. Here, $F_{0_\omega}[z,z']$ is the vacuum influence functional given as
\begin{equation}\label{1.39}
\begin{split}
F_{0_\omega}[z,z']=&\exp\biggr[-\frac{\mathcal{G}^2}{8\hbar m\omega}\int_0^Tdt\int_0^tdt' \left(X(t)-X'(t)\right)\\&\times\left(X(t')e^{-i\omega(t-t')}-X'(t')e^{i\omega(t-t')}\right)\biggr]~.
\end{split}
\end{equation}
This part of the calculation is done in the $c=1$ limit which will be restored later when we shall be calculating the uncertainty product. 
\begin{widetext}
\noindent The analytical form of $i\Phi_{\zeta_\omega}[z,z']$ is given by
\begin{equation}\label{1.40}
\begin{split}
i\Phi_{\zeta_\omega}[z,z']&=\frac{\mathcal{G}^2}{16\hbar m\omega}\int_0^T dt\int_0^T dt'\cos(\omega(t+t')-\phi_\omega)(X(t)-X'(t))(X(t')-X'(t'))\sinh 2r_\omega \\
&-\frac{\mathcal{G}^2}{16\hbar m\omega}\int_0^Tdt\int_0^Tdt'\cos(\omega(t-t'))
(X(t)-X'(t))(X(t')-X'(t'))(\cosh 2r_\omega -1)~.
\end{split}
\end{equation} 
If one now considers the real part of the squeezing parameter $r_\omega$ to be independent of the frequency \cite{QGravD} ($\phi_\omega$ is also replaced by $\phi$) and sum over all possible modes, the form of the total influence functional reads
\begin{equation}\label{1.41}
\begin{split}
F_\zeta[z,z']&=F_0[z,z']e^{i\Phi_z[z,z']}\\
&=\exp\biggr[-\frac{m_0^2G}{8\pi\hbar}\cosh 2r\int _0^\infty \omega d\omega\int_0^Tdt\int_0^Tdt'\cos(\omega(t-t'))(X(t)-X'(t))(X(t')-X'(t'))\\&+
\frac{m_0^2G}{8\pi\hbar}\sinh 2r\int_0^\infty\omega d\omega\int_0^Tdt\int_0^Tdt'\cos(\omega(t+t')-\phi)(X(t)-X'(t))(X'(t)-X'(t'))\biggr]\\&\times\exp\biggr[-\frac{im_0^2G}{8\pi\hbar}\int_0^Tdt (X(t)-X'(t))(\dot{X}(t)+\dot{X}'(t))\biggr]~.
\end{split}
\end{equation} 
\end{widetext}
It is important to note that the term involving the $\cos(\omega(t-t'))$ term has a time translational symmetry making it a static term whereas the term with $\cos(\omega(t+t')-\phi)$ breaks the time translational symmetry resulting in a non-static term. It is now possible to define two auxiliary functions given as
\begin{equation}\label{1.42}
\mathcal{A}^{s.}(t-t')=\frac{4\hbar G}{\pi}\cosh 2r\int_0^\infty d\omega\omega \cos(\omega(t-t'))
\end{equation}
\begin{equation}\label{1.43}
\mathcal{A}^{n.s.}(t+t')=\frac{4\hbar G}{\pi}\sinh 2r\int_0^\infty d\omega\omega \cos(\omega(t+t')-\phi)
\end{equation}
where `$s.$' and `$n.s.$' in the superscript of $\mathcal{A}$ denotes static and non-static respectively and we have set the $\phi\rightarrow 0$ limit. We shall be regularizing the integrals in eq.(s)(\ref{1.42},\ref{1.43}) by giving identical cut-off values to the upper limit and lower limit as before. If we denote the noise term corresponding to the static part as $\mathcal{N}^{s.}(t)$ then it is easy to identify $\mathcal{N}^{s.}(t)=\sqrt{\cosh 2r}\mathcal{N}_0(t)$ and the auxiliary function for the static part can be represented as $\mathcal{A}^{s.}(t-t')=\cosh2r\mathcal{A}_0(t-t')$. $\mathcal{A}_0(t-t')$ defined here is identical to $\mathcal{A}_0(t,t')$ defined earlier in eq.(\ref{1.25}). Hence, we can write down the transition probability as
\begin{widetext}
\begin{equation}\label{1.44}
\begin{split}
P^\Psi_{A\rightarrow B}\simeq& \int dz_i dz_i'dz_fdz_f'\phi_A(z_i)\phi^*_A(z_i')\phi_B^*(z_f)\phi_B(z_f')\int\tilde{\mathcal{D}}z\tilde{\mathcal{D}}z'\int\mathcal{D}\mathcal{N}_0\int\mathcal{D}\mathcal{N}^{n.s.} \exp\biggr[-\frac{1}{2}\int_0^Tdt\int_0^Tdt'\mathcal{A}^{-1}_0(t-t')\\\times&\mathcal{N}_0(t)\mathcal{N}_0(t')+\frac{1}{2}\int_0^Tdt\int_0^Tdt'{\left(\mathcal{A}^{n.s.}(t+t')\right)}^{-1}\mathcal{N}^{n.s.}(t)\mathcal{N}^{n.s.}(t')\biggr]\exp\biggr[\frac{im_0}{2\hbar}\int_0^Tdt\Bigr[(\dot{z}^2-\dot{z'}^2)-2(\phi-\phi')\\+&\frac{1}{2}(\sqrt{\cosh 2r}\mathcal{N}_0(t)+\mathcal{N}^{n.s.}(t))(X(t)-X'(t)\Bigr]-\frac{im_0^2G}{8\pi\hbar}\int_0^Tdt (X(t)-X'(t))(\dot{X}(t)+\dot{X}'(t))\biggr]
\end{split}
\end{equation}
where $\phi\equiv\phi(z)$ and $\phi'\equiv\phi(z')$.
\end{widetext}
Using the saddle point approximation,  we can now obtain the differential equation in $z$ as (neglecting higher order time derivatives terms of $z$)
\begin{equation}\label{1.45}
\begin{split}
\ddot{z}(t)+\frac{\partial\phi}{\partial z}-\frac{1}{2}\left(\ddot{\mathcal{N}}^{n.s.}(t)+\sqrt{\cosh 2r}\ddot{\mathcal{N}}_0(t)\right)z(t)=0~.
\end{split}
\end{equation}
As argued in \cite{QGravD}, we can always focus on the static part of the noise fluctuation and neglect the non-static part. Following the same calculation as observed in the vacuum state case, we can recalculate the square of variance in the position part as (restoring $c$ in the result)
\begin{equation}\label{1.46}
\begin{split}
(\Delta z)^2&\simeq \frac{4\hbar g^2 G}{\pi c^5}\cosh 2r\biggr[\frac{9z_0^2}{16\pi^2c^2}\left[\cos\left(\frac{2\pi c\tau}{z_0}\right]-1\right)\\&+\frac{1}{2}\tau^2\left(\text{Ci}(2\pi)-\text{Ci}\left(\frac{2\pi c\tau}{z_0}\right)\right)\\&+\frac{1}{8}\tau^2\ln\left(\frac{4\pi^2}{\tau^2}\right)+\tau^2\ln\left(\frac{c\tau}{z_0}\right)\biggr]\\
\implies (\Delta z)^2&\simeq \frac{4g^2\tau^2l_p^2}{\pi c^2}\cosh 2r\ln\left(\frac{c\tau}{z_0}\right)~.
\end{split}
\end{equation}
The square of the variance in the momentum parameter can be expressed as (keeping dominant terms only)
\begin{equation}\label{1.47}
(\Delta \pi_z)^2\simeq\frac{2\pi m_0^2g^2l_p^2\tau^2}{z_0^2}\cosh 2r~.
\end{equation} 
Using eq.(s)(\ref{1.46},\ref{1.47}), one can obtain the uncertainty product $\Delta z\Delta\pi_z$ as
\begin{equation}\label{1.48}
\begin{split}
\Delta z\Delta\pi_z&=\frac{2\sqrt{2}m_0g^2\tau^2l_p^2}{z_0 c}\cosh2r\sqrt{\ln \left(\frac{c\tau}{z_0}\right)}\\
&=\frac{2\pi m_0^2g^2\tau^2l_p^2}{z_0 c}\cosh2r\frac{z_0}{m_0 c}\sqrt{\frac{2}{\pi^2}\ln\left(\frac{c\tau}{z_0}\right)}\\
&=\frac{\beta z_0}{m_0 c}(\Delta \pi_z)^2
\end{split}
\end{equation}
which is identical to the result obtained in eq.(\ref{1.30}) for the vacuum state case. Following the arguments used in the vacuum state case, we can rewrite the uncertainty principle induced by the noise of gravitons as 
\begin{equation}\label{1.49}
\Delta z\Delta\pi_z\geq\frac{\hbar}{2}\left(1+\frac{\beta' l_p}{\hbar m_0 c}(\Delta \pi_z)^2\right)
\end{equation}
which again reduces to the well know generalized uncertainty principle for $m_0=m_p$. 

\noindent As before we can also introduce the same upper bound redefining a closed uncertainty relation given as
\begin{equation}\label{1.50}
\frac{m_p c}{\beta l_p}(\Delta z)^2>\Delta z\Delta \pi_z\geq \frac{\hbar}{2}\left(1+\frac{\beta' l_p}{\hbar m_0 c} (\Delta\pi_z)^2\right)~.
\end{equation}
In the next subsection, we shall continue this analysis for the graviton initially being in a thermal state. If we also obtain similar uncertainty relation, it will be a conclusive evidence of a new and universal uncertainty relation induced by the noise of gravitons.
\subsection{Thermal state}
\noindent For the gravitational fluctuation initially being in a thermal state, one can write down the influence functional  after summing over all possible frequency modes as \cite{QGravD}
\begin{equation}\label{1.51}
F_{th}[z,z']=F_0[z,z']e^{i\Phi_{th}[z,z']}
\end{equation}
where $i\Phi_{th}[z,z']$ (for $c=1$) is given as
\begin{equation}\label{1.52}
\begin{split}
i\Phi_{th}[z,z']&=-\frac{m_0^2G}{4\pi\hbar}\int_0^\infty\frac{\omega d\omega}{e^{\frac{\hbar\omega}{k_BT}}-1}\int_0^T\int_0^T dtdt'(X(t)\\&-X'(t))(X(t')-X'(t'))\cos(\omega(t-t'))~.
\end{split}
\end{equation}
The auxiliary function for such a thermal state can be written as 
\begin{equation}\label{1.53}
\begin{split}
\mathcal{A}_{th}(t,t')=\frac{8\hbar G}{\pi c^5}\int_0^\infty\frac{\omega d\omega}{e^{\frac{\hbar\omega}{k_BT}}-1}\cos(\omega(t-t'))
\end{split}
\end{equation}
where we have restored $c$ in the above equation. It is important to note that the auxiliary function is no more divergent in case of the gravitational fluctuation initially being in a thermal state. Hence, there is no need to regularize the upper and lower limits of integration as done for the vacuum and squeezed vacuum states.  The differential equation in $z$ (neglecting higher-order time derivative terms) is given as
\begin{equation}\label{1.54}
\ddot{z}(t)+\frac{\partial\phi}{\partial z}-\frac{1}{2}\left(\ddot{\mathcal{N}}_0(t)+\ddot{\mathcal{N}}_{th}(t)\right)z(t)=0~.
\end{equation}
We can mainly focus on the thermal part of the noise fluctuation. It is important to note that during the calculation of the $\llangle z_q(\tau)z_q(\tau)\rrangle$, several spurious terms occurred which have been neglected and only the dominant terms in the result have been used to obtain the uncertainty product. In this case, we can write down the square of the variance in the position part (keeping dominant contributions only)
\begin{equation}\label{1.55}
\begin{split}
(\Delta z)^2&\simeq\frac{2g^2\tau^2\hbar G}{\pi c^5}\ln\left[\sinh\left(\frac{\pi\tau k_B T}{\hbar}\right)\right]+\frac{5g^2G\tau^3k_BT}{ c^5}\\
&+\cdots\\
&\simeq \frac{2g^2\tau^2 l_p^2}{\pi c^2}\ln\left[\sinh\left(\frac{\pi\tau k_B T}{\hbar}\right)\right]
\end{split}
\end{equation}
where we have not considered the second term as it does not involve any $\hbar$ instead of being a term solely generated from quantum noise fluctuations. The square of the variance in the momentum part reads 
\begin{equation}\label{1.56}
\begin{split}
(\Delta\pi_z)^2&=\frac{\pi m_0^2g^2\tau^2 G k_B^2T^2}{3\hbar c^5}+\frac{2m_0^2\hbar Gg^2}{\pi c^5}\\&-\frac{2m_0^2k_BTGg^2\tau\coth\left(\frac{\pi
\tau k_BT}{\hbar}\right)}{c^5}\\&-\frac{2m_0^2g^2\hbar G}{\pi c^5}\ln\left[\frac{\pi
\tau k_BT}{\hbar}\text{cosech}\left(\frac{\pi
\tau k_BT}{\hbar}\right)\right]\\
\implies (\Delta\pi_z)^2&\simeq\frac{\pi m_0^2g^2\tau^2 G k_B^2T^2}{3\hbar c^5}=\frac{\pi m_0^2g^2\tau^2 T^2}{3T_p^2}
\end{split}
\end{equation}
where we have kept only the dominant terms in the final result and introduced the Planck temperature $T_p=\sqrt{\frac{\hbar c^5}{G k_B^2}}$.
Multiplying eq.(\ref{1.55}) with eq.(\ref{1.56}), we obtain the following relation
\begin{equation}\label{1.57}
(\Delta z)^2(\Delta \pi_z)^2=\frac{2m_0^2G^2k_B^2T^2g^4\tau^4}{3c^{10}}\ln\left[\sinh\left(\frac{\pi\tau k_B T}{\hbar}\right)\right]~.
\end{equation}
Hence, we can write down the following relation
\begin{equation}\label{1.58}
\begin{split}
\Delta z\Delta \pi_z&=\frac{\sqrt{2}m_0Gk_B T g^2 \tau^2}{\sqrt{3}c^5}\sqrt{\ln\left[\sinh\left(\frac{\pi\tau k_B T}{\hbar}\right)\right]}\\
&=\sqrt{\frac{6}{\pi^2}\ln\left[\sinh\left(\frac{\pi\tau k_B T}{\hbar}\right)\right]}\frac{\hbar}{m_0k_BT}(\Delta \pi_z)^2\\
\implies \Delta z\Delta\pi_z&=\frac{\gamma\hbar}{m_0k_BT}(\Delta\pi_z)^2
\end{split}
\end{equation}
where $\gamma=\sqrt{\frac{6}{\pi^2}\ln\left[\sinh\left(\frac{\pi\tau k_B T}{\hbar}\right)\right]}$~. The maximum possible value that $T$ can attain is the Planck temperature $T_p=\sqrt{\frac{\hbar c^5}{Gk_B^2}}$. In this limit, we can define an inequality involving the $\Delta z\Delta\pi_z$ product as
\begin{equation}\label{1.59}
\begin{split}
\Delta z\Delta\pi_z&=\frac{\gamma\hbar}{m_0k_BT}(\Delta\pi_z)^2\geq\frac{\gamma\hbar}{m_0k_BT_p}(\Delta\pi_z)^2\\
&=\frac{\gamma l_p}{m_0c}(\Delta\pi_z)^2~.
\end{split}
\end{equation} 
Following earlier arguments we can again write down the uncertainty relation induced by gravitons to be
\begin{equation}\label{1.60}
\Delta z\Delta\pi_z\geq \frac{\hbar}{2}\left(1+\frac{\gamma' l_p}{\hbar m_0c}(\Delta\pi_z)^2\right)
\end{equation}
with $\gamma'\equiv 2\gamma$.
As $\tau=\sqrt{\frac{2z_0}{g}}$, both $\gamma$ (for a fixed value of the temperature $T$) and $\beta$ are constants and it is always possible to set $\beta=\gamma$ (as a result $\beta'=\gamma'$). We can also recast eq.(\ref{1.58}) as
\begin{equation}\label{1.61}
\begin{split}
\Delta z\Delta \pi_z&=\frac{m_0k_BT}{\gamma\hbar}(\Delta z)^2\leq \frac{m_0k_BT_p}{\gamma\hbar}(\Delta z)^2\\
&=\frac{m_0 c}{\gamma l_p}(\Delta z)^2\leq\frac{m_p c}{\gamma l_p}(\Delta z)^2 
\end{split}
\end{equation}
Using eq.(\ref{1.60}) along with eq.(\ref{1.61}), we can write down the closed form of the uncertainty relation
\begin{equation}\label{1.62}
\frac{m_p c}{\gamma l_p}(\Delta z)^2> \Delta z\Delta \pi_z\geq \frac{\hbar}{2}\left(1+\frac{\gamma' l_p}{\hbar m_0c}(\Delta\pi_z)^2\right)~.
\end{equation}
For $\gamma=\beta$, we get back the same inequalities as obtained in eq.(s)(\ref{1.37},\ref{1.50}). Eq.(\ref{1.60}) again reduces to the usual generalized uncertainty relation in the $m=m_p$ limit. It is important to note that for the gravitational fluctuation being in a vacuum state, squeezed vacuum state, and a thermal state, we obtain the same uncertainty relation as can be observed from eq.(s)(\ref{1.37},\ref{1.50},\ref{1.62}).

\noindent The only limitation of this new uncertainty product obtained in eq.(s)(\ref{1.34},\ref{1.49},\ref{1.60}) is the involvement of a mass term in the denominator of the correction term in the lower bound which prevents it from being applicable to massless particles. Here, $m_0$ is the mass of the particle for velocity $\dot{z}(t)\ll c$, therefore we can always consider it to be the rest mass of the particle. Therefore, for a massless particle, we can replace $m_0$ by $\frac{\hbar \omega_0}{c^2}$ with $\omega_0$ denoting the frequency associated with the massless particle.
Hence, for a massless particle the uncertainty relation reads
\begin{equation}\label{1.63}
\frac{c^3}{\beta G}(\Delta z)^2>\Delta z\Delta\pi_z\geq \frac{\hbar}{2}\left(1+\frac{\beta' l_p c}{\hbar^2\omega_0}(\Delta\pi_z)^2\right)~.
\end{equation}
\section{Conclusion}\label{S4}
\noindent In this paper, we have extended the calculations presented in \cite{AppleParikh} and extended the analysis in order to obtain an uncertainty relation induced by the noise of gravitons. Contrary to the calculations presented in \cite{AppleParikh}, we have restricted our calculations to Newtonian corrections only as the post-Newtonian corrections do not have any direct influence on the noise part of the modified Langevin equation. In \cite{AppleParikh}, the variance in the position was calculated where the free fall of a point particle under the effect of Earth's gravity from a certain initial height was considered. The standard deviation in position was found to be dependent on the square root of a two point correlation function for a stochastic noise term. The stochastic noise term depends on the nature of the quantized gravitational field. In our analysis, we have progressed a step further by calculating the variance in the momentum of the point particle at the moment when the point particle touches the surface of Earth. We have found out an exact expression for the product of the position and momentum uncertainties for any point particle with mass $m_0$. We then have obtained a lower bound to the uncertainty product which depends on $\hbar$, $G$, $c$, and the square of the variance in the momentum variable. To our surprise, after combining this product with the standard Heisenberg uncertainty principle \cite{SCARDIGLI}, we obtain in the Planck mass limit the well known form of the uncertainty relation obtained in  \cite{Kempf}. We have calculated the analytical form of the $\Delta z\Delta\pi_z$ product term in the cases when the gravitational fluctuation is initially in a vacuum, squeezed vacuum, and thermal state. For the gravitational wave to be in a vacuum state, we observe that the product of the variance in the position and the variance in the momentum term is proportional to the square of the variance in the momentum term multiplied by $\frac{\beta z_0}{m_0 c}$ term. Here, $\beta$ is a dimensionless parameter and $z_0$ is the initial height of the freefall. Using the fact that the lowest value one can go to in a quantum gravity setting is the Planck length, we obtained the lower bound of the uncertainty product by summing the usual uncertainty product with this new product where we have replaced $z_0$ by the Planck length. This setting can physically be interpreted by the freefall of a massive particle from the height which equals the Planck length. Although it is practically impossible to do such an experiment, the limit is imposed only to obtain a lower bound on such an uncertainty product. It is now important to note that a further lower limit can be obtained if the mass of the particle is replaced by the Planck mass which gives us the usual generalized uncertainty principle with quadratic order correction in the momentum parameter \cite{Kempf}. Next, we observe that the $\Delta z\Delta\pi_z$ product can also be represented in terms of the square of the variance in the position parameter, and replacing $m_0$ and $z_0$ by their respective Planckian counterparts, we obtain a strong upper bound on the uncertainty product. The existence of a $(\Delta z)^2$ dependent upper bound suggests that the position of the particle can never be measured so precisely that $\Delta z$ turns out to be zero. It will always have a nonzero value but can become infinite if the momentum of the particle is measured precisely. Hence, from the total uncertainty relation, we interpret that $\Delta z$ can never be zero and as a result, $\Delta \pi_z$ can never be infinity. For the next part of our calculation, we consider the initial gravitational fluctuation to be in a squeezed vacuum state. This analysis indicates that for a squeezed state with a very high value of the squeezing parameter $r$, both $\Delta z$ and $\Delta \pi_z$ get exponentially enhanced by a factor $\sqrt{\cosh 2r}\sim e^r$ but the total uncertainty product attains the same value as for the case of the vacuum state analysis. We also obtain the same upper and lower bounds for the uncertainty product for the gravitational fluctuation being in a squeezed vacuum state initially. Finally, we extend our calculation for the gravitational fluctuation being initially in a thermal state. Contrary to the other cases, the auxiliary function does not diverge in case of the gravitational fluctuation being in a thermal state and as a result, one need not regularize the two limits of integration in the frequency variable. As before, replacing the temperature of the thermal gravitational interaction with that of the Planck temperature, we get back the earlier upper and lower bounds. The three examples solidify the new form of the graviton-induced uncertainty principle with an overall mass dependence in the correction term. It is interesting to note that the usual form of the generalized uncertainty product used in most of the literature \cite{OTMGraviton,Kempf,MAGGIORE,SCARDIGLI,ADLERSAN,
ADLERCHENSAN,RABIN,SG1,SCARDIGLI2,SG2,Ong,EPJC,BMajumder,DAS1,DAS2,
IVASP,IVASP2,KSP,SG3,SG4,OTM,Petruzzeillo,
DasModak,Farag,Farag2,SGSB,OTM0,Vagnozzi,Feng}  is completely $\hbar$ independent, which is very unusual for a quantum gravity correction. A true quantum gravity correction must involve both Planck's constant and Newton's gravitational constant. Our analysis produces a term that is truly quantum gravitational in nature as it involves both $\hbar$ and $G$. The reason behind the appearance of such a term is solely due to the interaction of the particle with the graviton. It is also important to note that in the case of the particle mass being replaced by the Planck mass, we obtain the usual generalized uncertainty relation indicating a derivation of the generalized uncertainty principle from the noise induced by gravitons.  Finally, we would like to point out that it is quite remarkable that the Langevin equation for the point particle falling in the gravitational field of the Earth gives rise to a stochastic uncertainty principle which is consistent with the generalized Heisenberg uncertainty principle. The reason behind this consistency is due to the coupling of classical degrees of freedom corresponding to the particle with the quantized gravitational field\footnote{This was pointed to us by Maulik Parikh.}.
\section*{Acknowledgement}
\noindent We thank the referee for useful comments on our paper. We also thank Dr. Maulik Parikh \textit{(Professor, Department of Physics and The Beyond Center, Arizona State University)} for an email correspondence over the manuscript.  

\end{document}